\pdfoutput=1
\documentclass[showpacs,twocolumn,prl]{revtex4}
\usepackage{epsfig}
\usepackage{amsmath}
\usepackage{graphicx}
\usepackage{amssymb} 
\usepackage{graphics}

\begin{document}

\title{Nanoscale Enzyme Field Effect Transistor for Urea Sensing }

\author{Yu Chen, Xihua Wang, Mi Hong, Shyamsunder Erramilli, Pritiraj Mohanty}

\affiliation{
Department of Physics, Boston University, 590 Commonwealth Avenue, Boston, MA 02215}

\date{\today}

\begin{abstract}

Silicon nanowires have been surface functionalized with the enzyme urease for biosensor applications to detect and quantify urea concentration. The device is nanofabricated from a silicon-on-insulator (SOI) wafer with a top down lithography approach. The differential conductance of silicon nanowires can be tuned for optimum performance using the source drain bias voltage, and is sensitive to urea at low concentration. The experimental results show a linear relationship between surface potential change and urea concentration in the range of 0.1-0.68 mM. The sensitivity of our devices shows high reproducibility with time and different measurement conditions. The nanowire urea biosensor offers the possibility of high quality, reusable enzyme sensor array integration with silicon-based circuits.

\end{abstract}


\maketitle

\textbf{I. Introduction:}\\
\\
Since the concept of ISFET (ion sensitive field-effect transistor) was first introduced in biosensor applications by Bergveld\cite {Berg70}, it has attracted a lot of interest among both experimentalists and theorists\cite{Torb06,Vumm05,Will00,Shim07,Li00}. The surface of the gate of the FET device can be modified with sensing molecules like antibody or antigen, and has the potential for serving as a highly efficient immunological sensor with the required specificity and sensitivity\cite{Jun07}. When antigen or antibody in the test solution reacts with the sensing molecules on the surface, any change in the charge state or surface potential leads to modulations of the conductance channel of the FET device. Thus the concentration of the test molecules can be extracted by the conductance or current measurement of the device. The measurement of the immunological reaction is limited by the sensor response and also ionic strength and pH of the test sample\cite{Berg03}. 
	
For potential clinical applications, an enzyme FET (ENFET), which is a pH-based ISFET has been developed\cite{Berg03,Shim07}. Glucose sensors, and urea sensors are among the best-known biosensors for widespread clinical applications. Traditional enzyme-linked immunosorbent assay (ELISA)\cite{Leik01} measure the spectrum of the enzyme-catalyzed products and require large volume of the sample solution and thus decrease the sensitivity of the device. The ENFET, especially micro and nanoscale FET, does not need large volume of the solution and can provide fast and high sensitive detection of target molecules.

Semiconductor nano-channels, such as carbon nanotubes (CNTs) and silicon nanowires (SiNWs), have been studied intensely for their extraordinary electrical, mechanical and optical characteristics\cite{Qin07,Dekk98,Will04}. These characteristics have been used for ultrasensitive biosensor applications [8] and [14]. Taking the advantage of large surface-to-volume ratio, their electronic conductance may even be sensitive enough to detect single molecule binding to the surface. Most of the existing studies based on a bottom-up fabrication approach are difficult to integrate into the manufacture of complex sensing circuits. Fabrication of silicon nanowires with the top down approach\cite{Moha06,Reed07}, based on standard semiconductor processes, offer the promise of manufacturability and scalability for mass production. Thus, field effect devices, combining nanotechnology, offer the possibility to produce high-performance, low-cost biosensors.

\begin{figure} [t]
	\includegraphics[scale=0.45]{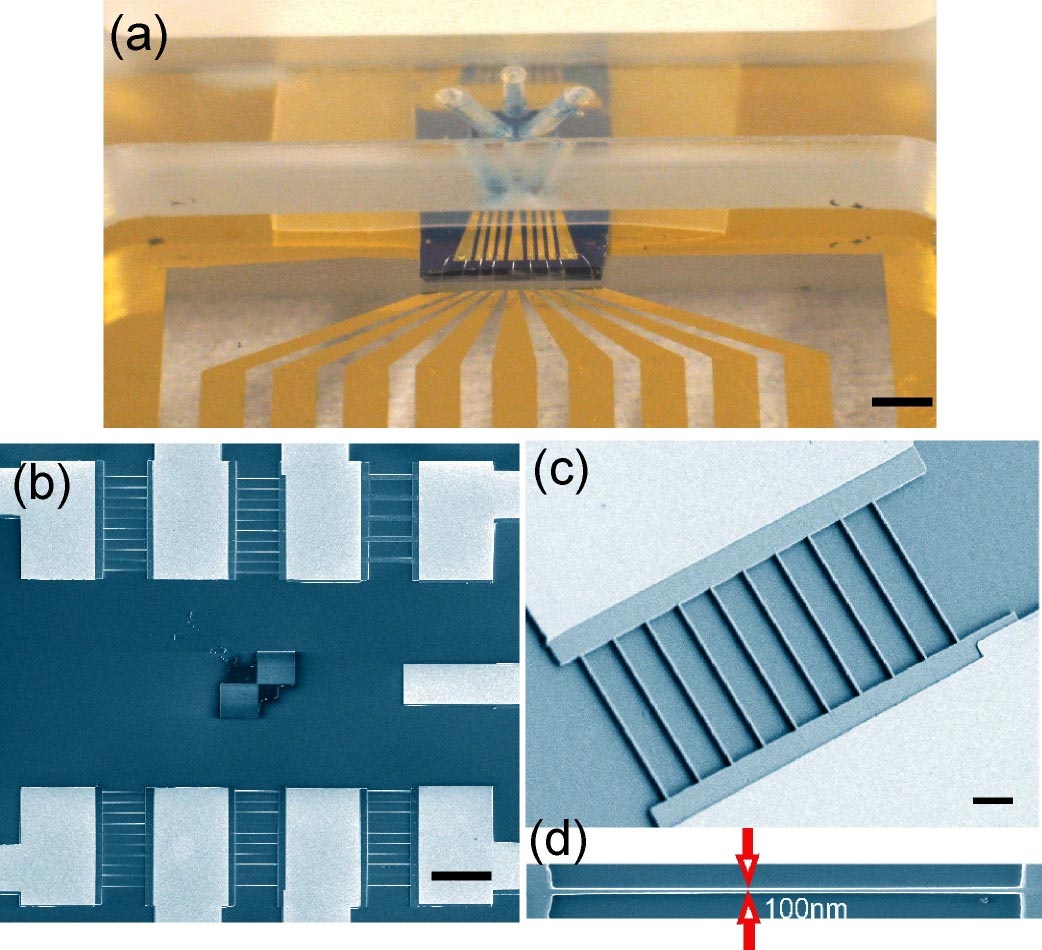}
	\caption{ (a) Optical micrograph of the device, integrated with the solution exchanging chamber. (b), (c), (d) are scanning electron micrographs of the sample with six devices integrated in a single chip, a single device and a single wire. The scale bars are 1mm, 10$\mu$m, 2$\mu$m in (a) (b) and (c).}
	\label{fig1}
\end{figure}

Here we demonstrate the silicon nanowires, surface functionalized with enzyme (urease), as a field-effect enzyme biosensor. In the differential conductance measurements, the device shows high sensitivity to the local change in hydrogen ion concentration produced by the enzymatic reaction, i.e. essentially a "local" pH change. We note that pH is usually defined as an inherently equilibrium concept. The FET sensor senses the change in the surface potential due to a change in concentration of hydrogen ions near the surface of the sensor. To the extent that the time scales of measurement are kept slow compared to the time scale of exchange of hydrogen ions, the surface potential can still be linked to an effective pH. The sensitivity of the device response to the urea is demonstrated down to 0.1 mM. While the device can be made sensitive to still lower concentrations, this limit is sufficient for the clinical applications, which require operation in the 0.1-1 mM range. The calibrated surface potential change introduced by the reaction has a linear range for urea concentration between 0.1 mM and 0.68 mM. Our silicon wire FET sensor shows very good stability. The dependence of the differential conductance on urea concentration varies with the drain voltage $V_{ds}$. However, the dependence of the surface potential on the urea concentration is independent of $V_{ds}$. This provides a simple method of calibrating the response.\\

\textbf{II. Experimental methods:}\\
(a). {\it Device fabrication:} The urea biosensor with a set of silicon nanowires is fabricated from silicon-on-insulator (SOI) wafer. The SOI wafer consists of 100 nm thick silicon as a device layer, 380 nm $SiO_{2}$ as an insulation layer and a 500 米m thick silicon substrate. The silicon device layer is lightly doped with boron concentration $1~2\times10^{-15} cm^{-3}$. The silicon wires are typically 4 to8 $\mu$ m long, and the width can be designed down to 50 nm in our experiments. The silicon nanowires are first patterned with electron beam lithography. Then a not, vert, similar40 nm layer of chromium is evaporated as a mask by thermal evaporator. Further reactive ion etching (RIE) exposes the two side walls of silicon wires. Source and drain contacts are also defined by electron beam lithography and Ti/Au are deposited in a thermal evaporator, without further high temperature annealing process or doping. A typical chip in our experiments includes six devices. Fig. ~\ref{fig1}(b) shows a scanning electron micrograph of the device with six devices in a single chip. Fig.~\ref{fig1}(c) is a single device with multiple wires. Multiple wires design increases the measurement signal (conductance of the device) while keeping the small surface-to-volume ratio, so the signal noise ratio. Fig. ~\ref{fig1}(d) is a single wire with 100 nm width. The silicon nanowires were further covered with 10 nm Al2O3, grown by atomic layer deposition (ALD), to prevent current leakage between analyte solution and silicon nanowires

\begin{figure} [t]

	\includegraphics[scale=0.15]{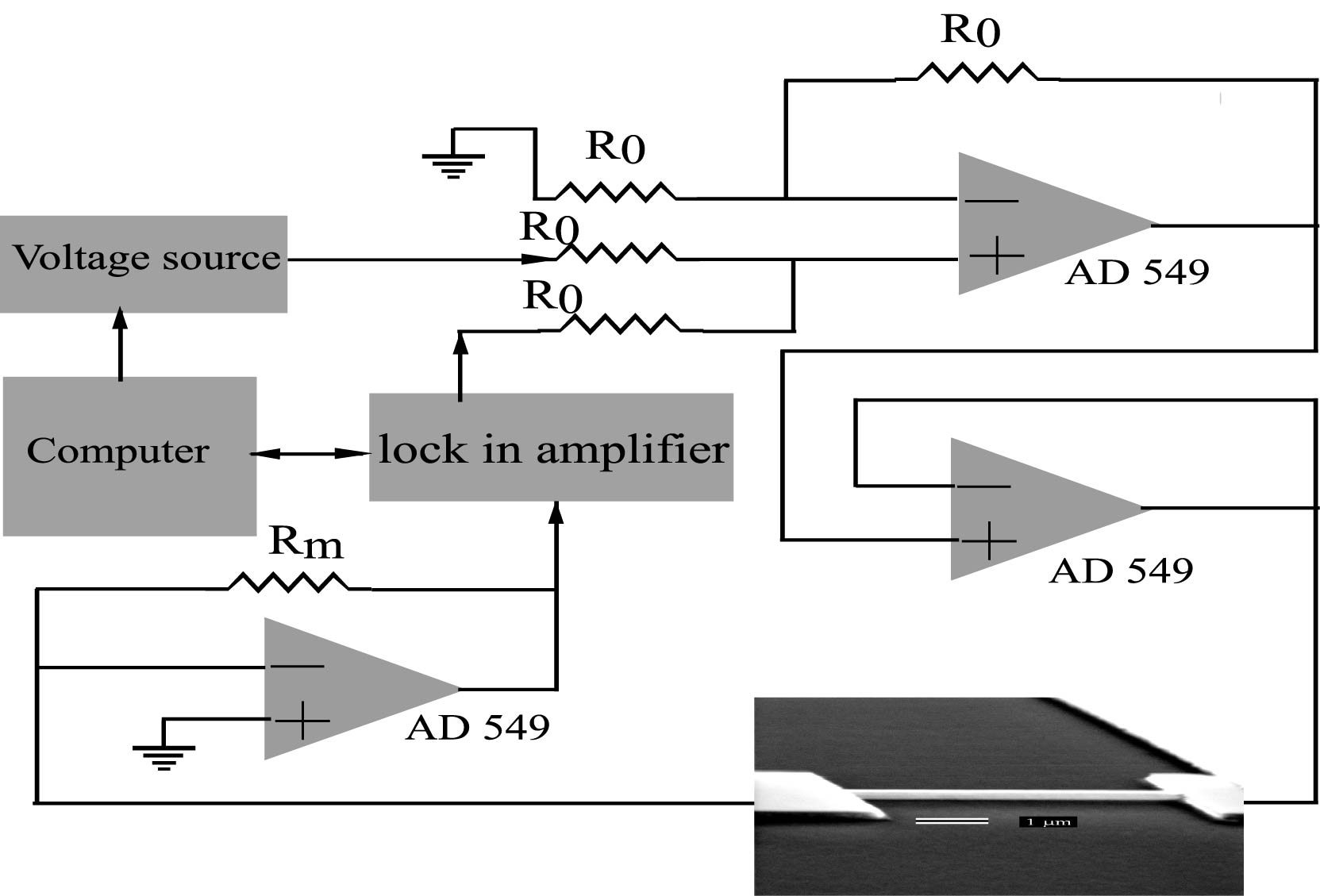}
	
	\caption{Differential conductance measurement circuit. PLEASE EXPAND ON THE CAPTION SOMEWHAT.}
	
	\label{fig2}

\end{figure}

(b). {\it Surface modification:} The silicon wire FET sensor is modified with urease following procedure below. Before modification, the Al$_2$O$_3$ surface was treated with oxygen plasma [14] (50 mW power, 30 sccm flow rate for 1 min) for two purposes. One is for cleaning the sample surface, and the other one is for generating a hydrophilic surface. The wires are first put into 3-aminopropyltriethoxy silane (APTES) solution (3\% in ethanol with 5\% water) for 2 h. The device is rinsed with ethanol solution for five times before baking at $110 ^o$C for 10 min. After wire bonding, 2 urease in 20 mM NaCl solution (5\% glycerol, 5\% Bovine Serum Albumin, BSA) is deposited on the sample and is kept in glutaraldehyde vapour for 40 min. The sample is then dried in air for 15 min. After modification, the sample is covered with solution-exchanging chamber and the devices are kept in buffer solution before further calibration and measurements. All urea samples used in our experiment contain 50 mM NaCl solution.

(c). {\it Solution exchanging chamber:} The solution-exchanging chamber is made of polycarbonate (PC), with three holes (1.1 mm diameter), one serving as an inlet channel, a second as a fluid output channel and the third for insertion of a reference gate electrode. Fig. ~\ref{fig1}(a) shows a chip placed under a solution-exchanging chamber. A thin parafilm spacer with a hole in the centre is inserted between the chamber and the chip to enable a small volume of solution to flow across on the device. The typical volume of the solution used for the experiments is 20 to 30 $\mu$l. This configuration offers superior time response when compared to microfluidic chambers where laminar flow can dominate.

(d). {\it Measurements setup:} The measurement circuit includes a small ac modulation (provided by an EG\&G 5210 lock-in amplifier), superimposed on the dc bias across the nanowire (provided by a Keithley 2400 source meter). The ac modulation and the dc bias are added by a non-inverting summing circuit, integrated with the preamplifier circuit (Fig. ~\ref{fig2}). The entire device is placed in an RF-shielded aluminium box to prevent noise pickup. Differential conductance measurements are done by sweeping the dc bias at constant ac modulation amplitude, and measuring the response with the lock-in amplifier, referenced to the ac signal frequency. The quantity of interest is the change in $g$, the differential conductance due either to a change in the reference gate voltage $V_{rg}$, or to a change in concentration $C$.\\

\begin{figure} [t]
	
	\includegraphics[scale=0.4]{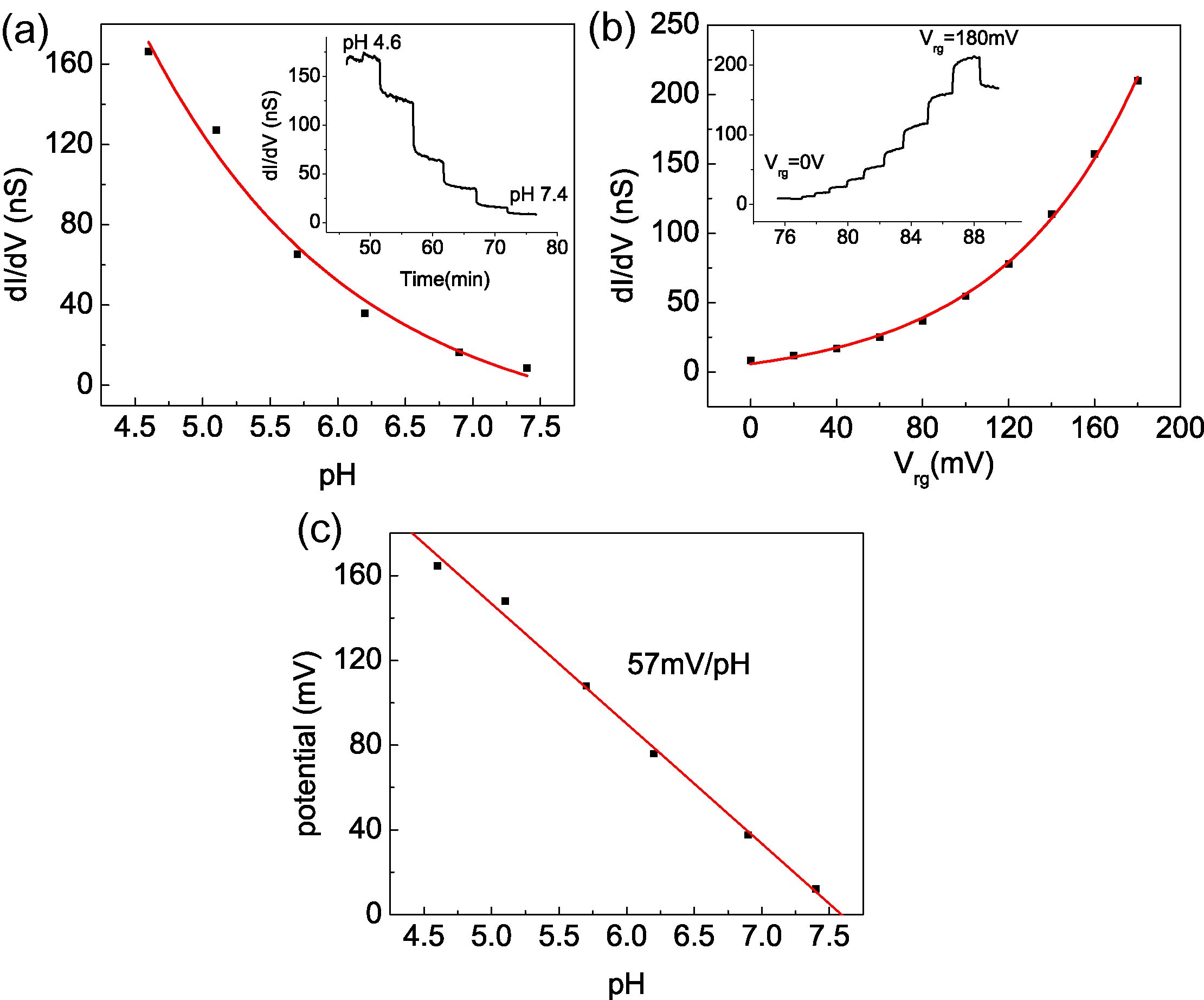}
	
	\caption{Sensor sensitivity calibration with pH measurements. (a) Real time differential conductance measurement when reference gate voltage is changed sequentially while keeping the device in same solution. $V_{ds}=0$ V (b) Real time differential conductance measurement when the pH value of the solution is changed sequentially while the reference gate voltage is fixed. $V_{ds}=0$ V, $V_{rg}=0$ V. (c) Differential conductance versus pH value of the solution and reference gate voltage. 	}
	
	\label{fig3}
	
\end{figure}

\textbf{III. Results and discussion:}\\
\\Urea is also known as carbamide and it was the first organic compound to be artificially synthesized from inorganic starting materials [17]. The monitoring of urea concentration in blood is a way to evaluate kidney disease\cite{Yild05}.

When urea reaches the functionalized surface, the enzyme catalyzes the following reaction\cite{Toum07}:
\begin{equation}
	\mbox{Urea}+3H_{2}O\stackrel{\mbox{Urease}}{\rightarrow}2NH_4^{+}+HCO_3^{-}+OH^-
\end{equation}
\begin{figure} [t]
	\includegraphics[scale=0.13]{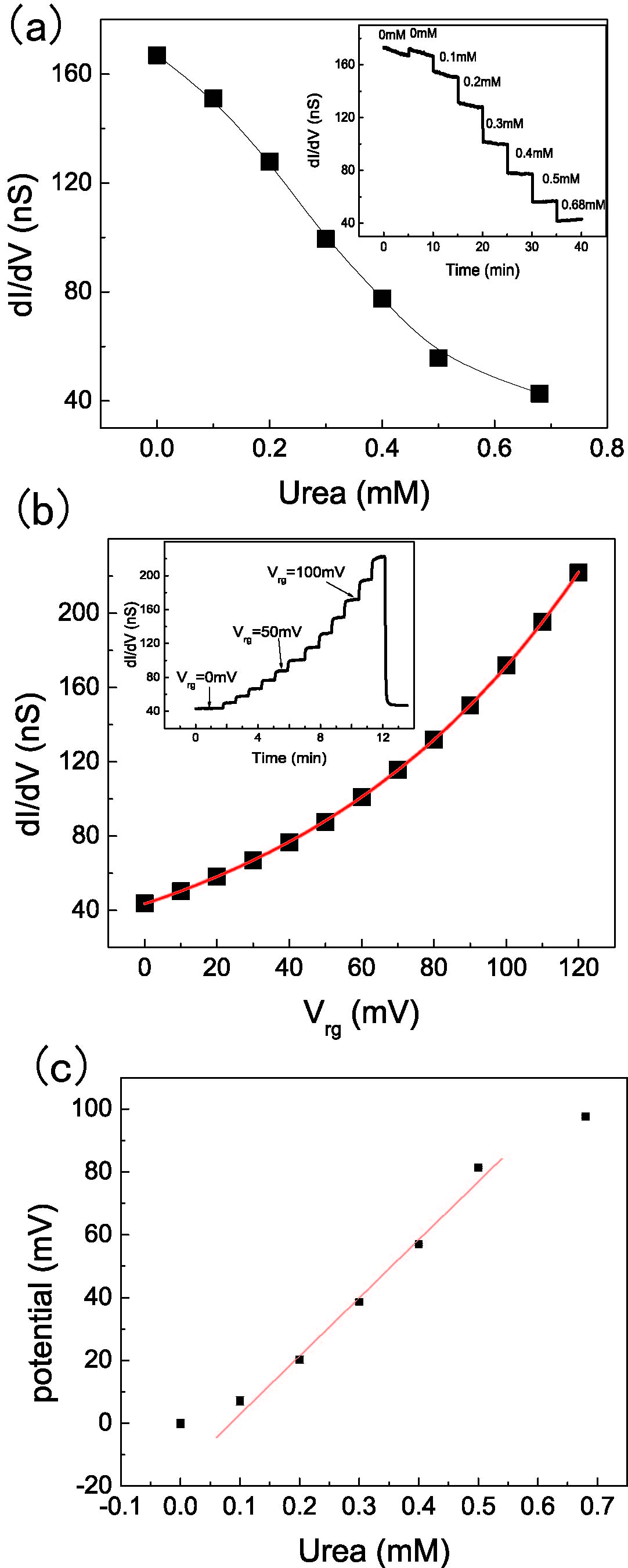}
	\caption{(a) Differential conductance versus urea concentration, Vds=-0.6V, Vrg=0V. Inset is real time differential conductance measurement of the device when the urea concentration is changed, all solutions contents 50mM of NaCl (b) Differential conductance versus reference gate voltage in same solution. Inset is the real time differential conductance measurement. (c) Surface potential change versus urea concentration. Data is normalized from (a) and (b).}
	\label{fig4}
\end{figure}

Although both $NH_4OH$ and $H_2CO_3$ affect the hydrogen ion concentration, and hence the pH, of the solution, the dissociation constant$K_D$ of $NH_{4}OH$ is $1.8\times10^{-5}$ M,  while $K_{D}'$ of $CO_{2}$ is $3\times10^{-7}$M,  the $NH_4OH$ is stronger than carbonic acid created by $CO_2$, resulting in an increase in the pH at the gate interface. Roughly speaking, the enzyme FET sensor may be considered to function as a pH sensor. The sensitivity of the device depends on the sensitivity to the local pH change of the solution introduced by the catalytic reaction. We first characterize the pH sensitivity of the device with silicon wires without enzyme modification. Fig. ~\ref{fig3}(a) shows the relationship between the differential conductance change and pH value of the solution. The inset is the real time measurement of the differential conductance of the device, the wire width is 100 nm and covered with 10 nm of Al$_2$O$_3$, the reference gate voltage is set to be 0V. When the pH value of the solution decreased, surface potential $\psi$ on the silicon wire increased\cite{Berg03}:

\begin{equation}
\Delta\Psi=-2.3\alpha\frac{kT}{e}\Delta pH
\end{equation}

\begin{figure} [t]
	\includegraphics[scale=0.13]{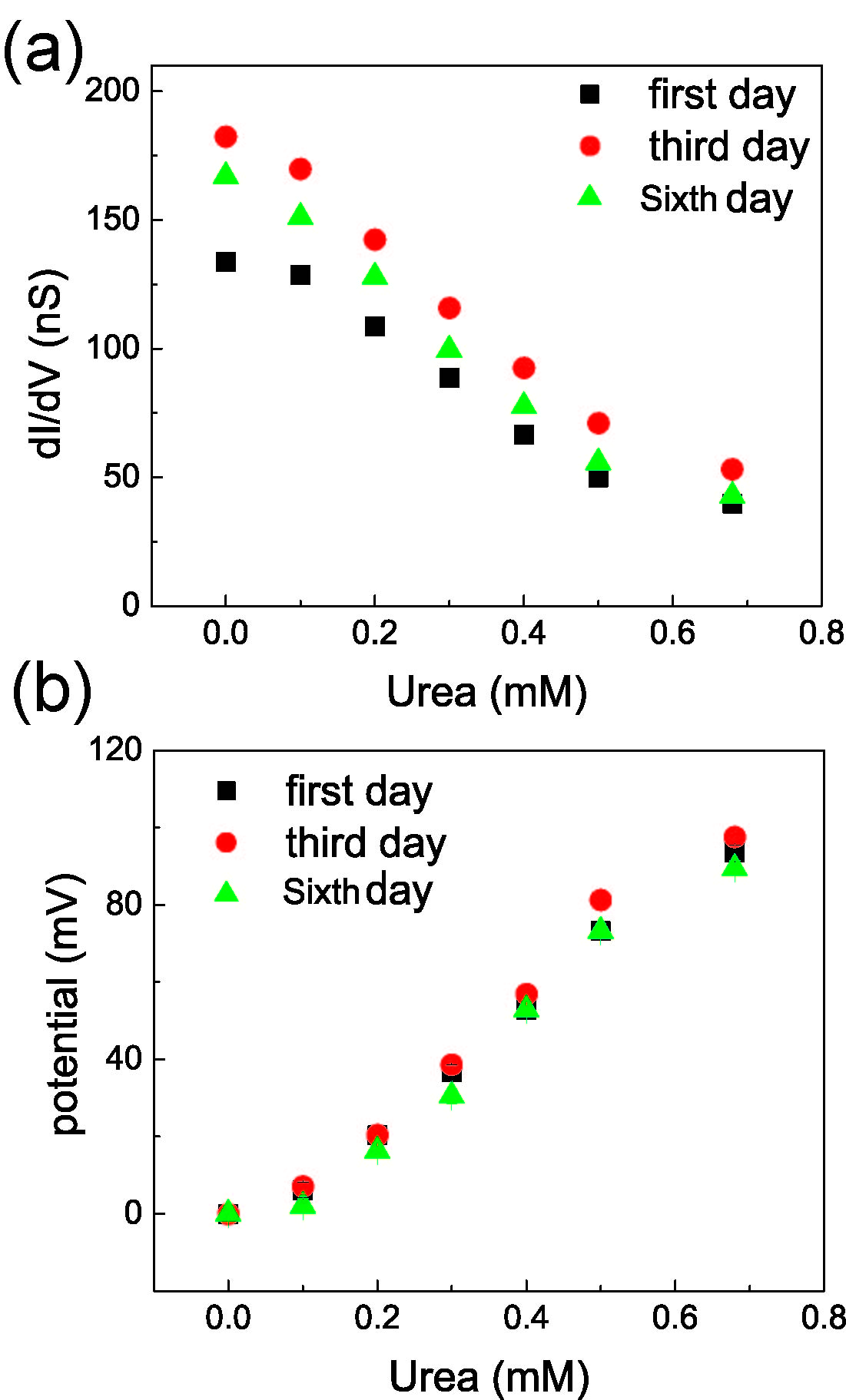}
	\caption{ Stability of the device (a) Differential conductance versus urea concentration, data are taken during six days after device is ready. (b) Surface potential change versus urea concentration, data is normalized by reference gate voltage change.  }
	\label{fig5}
\end{figure}
q is the proton charge, k is Boltzmann's constant, and T the absolute temperature. $\alpha$ is constant related to the buffer capacity of the surface and it is between 0 and 1. The physical quantity we are measuring in our experiments is device conductance. In order to calibrate the pH change to conductance change of the device, we change the reference gate voltage while keeping the solution character the same. For our device, as can be seen in Fig. ~\ref{fig3}(b), conductance is not in a linear relation with gate voltage. This is because the existence of the Schottky barrier of the two electrical contacts. By operating the device at small negative source drain voltage, the device can be used as ion sensitive sensor. By converting the conductance change introduced by urea to the reference gate voltage change which is proportional to the surface potential change in Fig.~\ref{fig3}(c), we find the sensitivity is around 57 mV/pH, which is close to the limit of most FET sensors [8]. This calibration confirmed that nanoscale FET sensor has comparable if not better sensitivity to pH comparing with large-scale FET sensor and shows very high stability at the same time.

\begin{figure} [t]
	\includegraphics[scale=0.6]{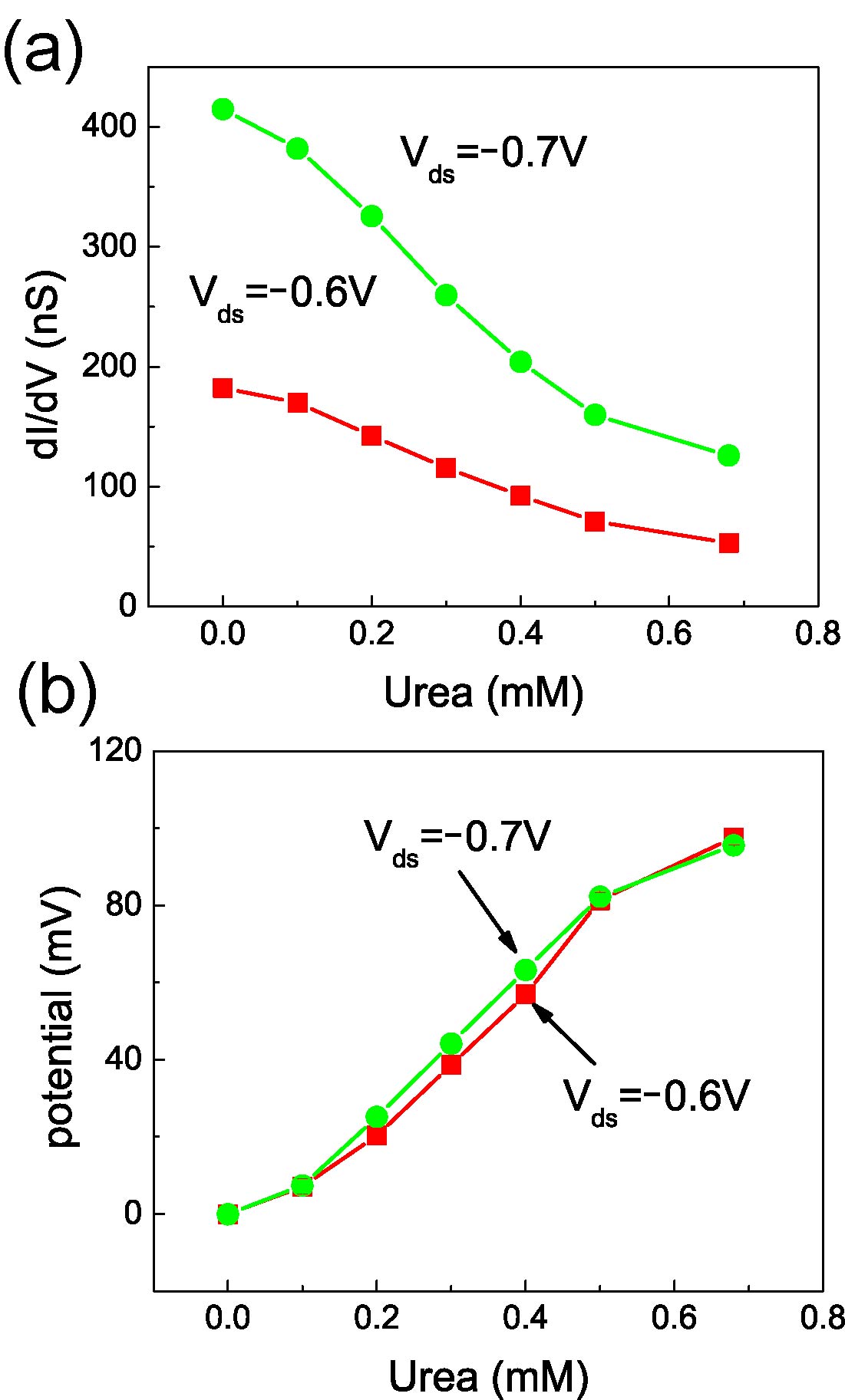}
	\caption{ (a) Differential conductance versus urea concentration when data are taken different source drain voltage: $V_{ds}= 0.6$ V and $V_{ds}= 0.7$ V. (b) Surface potential change at differential source drain voltage. Data is normalized by reference gate voltage change.}
	\label{fig6}
\end{figure}

Fig.~\ref{fig4} gives the real time response of the device surface modified by urease enzyme when adding urea solutions with different concentrations. After each solution change the data taking is paused for 10 min to wait for transient signals to decay and the response to stabilize. The time response is limited not by the diffusion time scale over the small sample chamber but rather by the presence of multiple layers of enzyme molecules on the surface. Further improvement of the response time can be done by modifying the surface with monolayer enzyme\cite{Khar00}. After urea measurement, the reference gate voltage is changed to get a relationship between the gate voltage and conductance similar to what we did in pH measurements. By comparing the data during urea concentration change and during the reference gate voltage change, we can normalize the urea concentration to reference gate voltage (surface potential) change [20]. Fig. ~\ref{fig4}(c) shows a monotonic dependence of the surface potential on the urea concentration, with a linear dependence in the urea concentration range from 0.1-0.68 mM. When increasing the urea concentration, the results saturates because of the activity change of the urease enzyme. From our data, our devices are useful in a high sensitive sensing application.

In order to satisfy clinic applications, such as long time monitoring of the target concentration and reusable sensor, it is important to have highly stable sensor. Further experiments are done to study the stability of our sensor by measuring the conductance change due to the presence of urea. Fig. ~\ref{fig5}(a) shows the conductance change at first day, third day and sixth day. Like many silicon devices, due to drifting problem of the silicon device we noticed that the conductance change can be up to 25\% (compare the conductance change introduced by 0.68 mM urea of first day and third day). But if we normalize the data with surface potential change by doing reference gate measurement like in Fig.~\ref{fig5}(b), we found that the drifting potential is less than 10\%, which offers possibility of high repeatable and reusable urea sensors.

The surface potential change introduced by the urea concentration is only related to the solution character and should not depend on the sensor device or on the measurement conditions. In order to confirm this we did the same measurements by change the source drain bias voltage. Fig. 6 shows the data taken at $V_{ds}=-0.6$ V and $V_{ds}=-0.7$ V. As described in a previous paper\cite{Roth07}, our device response gives bigger signal when $V_{ds}$ is set at higher negative value; this is consistent with the data in Fig. 6(a). But if we normalize the conductance change to the surface potential change of the device like in Fig.~\ref{fig4}, the data in Fig.~\ref{fig1}(b) shows that the surface potential change due to the catalytic reaction is not affected by the source drain voltage.\\

\textbf{IV. Conclusion:}\\
\\
Silicon nano-channel FET sensor is fabricated from a top down lithography approach. The device shows a good sensitivity to the pH change of the solution. This confirmed that the surface potential change of the nanoscale device introduced by the pH value of the solution is comparable to if not better than large-scale devices. The surface of the silicon-channels is further modified with urease enzyme and used for the detection of the concentration of the urea. The device response is in a linear relationship with the urea concentration of the solution in the range of 0.1每0.68 mM. Our nanoscale device shows very good stability and offers the possibility of highly efficient, repeatable enzyme sensor array.


\end{document}